\input BoxedEPS
\SetTexturesEPSFSpecial

\documentstyle[prl,twocolumn,aps]{revtex}

\begin{document}

\draft
\twocolumn[\hsize\textwidth\columnwidth\hsize\csname @twocolumnfalse\endcsname
\title{Aharonov-Bohm cages in two-dimensional structures}
\author{Julien Vidal$^{1}$, R\'{e}my Mosseri$^{1}$ and Benoit Dou\c{c}ot$^{2}$}
\address{$^{1}${\it Groupe de Physique des Solides, }\\
$^{2}${\it Laboratoire de Physique Th\'{e}orique des Hautes Energies,}\\
{\it \ Universit\'{e}s P. et M. Curie P6 et D. Diderot P7, Tour 23, 2 place Jussieu, 75251 Paris}
Cedex 05\\}
\maketitle

\begin{abstract}
We present an extreme localization mechanism induced by a magnetic field for tight-binding electrons in two-dimensional
structures. This spectacular phenomenon is investigated for a large class of tilings (periodic, quasiperiodic, or
random). We are led to introduce the Aharonov-Bohm cages defined as the set of sites eventually
 visited by a wavepacket that can, for particular values of the magnetic flux, be bounded. We finally discuss
the quantum dynamics which exhibits an original pulsating behaviour.
 
\end{abstract}

\pacs{PACS numbers: 7100, 7210, 7335.}
\vskip2pc]

\vspace{.5cm}

Over the past two decades, the behaviour of electrons in two-dimensional structures under a magnetic field
has been of special interest in condensed matter physics. In situations where electron-electron interactions
are dominant (low carrier density heterostructures for instance), a strong magnetic field has been shown to induce
new collective states such as fractional quantum Hall states \cite{a} or Wigner crystals \cite{b}. But even when electrons
 may be considered as weakly interacting, fascinating phenomena may take place in the presence of a 
spatially modulated potential besides the external magnetic field. In the case of a random potential,
 one obtains the integer quantum Hall effect \cite{c} which is characterized by a complex one-particle spectrum of 
localized states and a discrete set of eigenenergies associated to extended states \cite{d}. If the modulated 
potential corresponds to a periodic lattice, a subtle competition between this periodicity and the area
scale imposed by the magnetic field induces  remarkable features for the one-particle spectrum \cite{Azbel},
illustrated by the well-known Hofstadter's butterfly-like pattern\cite{Hof76}.
The physics of this magnetically induced frustration has also generated many
experiments on submicronic structures such as superconducting networks \cite{Pannetier}. They have in 
turn triggered investigations of several different two-dimensional structures, like the triangular lattice \cite{Claro79},
the honeycomb tiling \cite{Rammal}, the quasiperiodic tilings\cite{Schwabe}, and even fractal networks 
\cite{Doucot} \cite{e}.
\\
In this letter, we discuss a new effect of extreme localization in a large class of rhombus tilings. 
For particular values of the magnetic field, the set of sites visited by an initially localized
 wavepacket can be bounded 
due to Aharonov-Bohm ($\cal{A.B.}$) destructive interferences\cite{Bohm}. We are thus led to introduce the notion of 
$\cal{A.B.}$ cages. Our aim is to analyze the relationship between this phenomenon, the tiling geometry, the quantum dynamics, and the
 spectrum. We therefore pay attention to a periodic tiling with hexagonal symmetry \cite{Sutherland},
 here denoted  ${\cal T}_{3}$  (see figure \ref{Star3lattice}). We show that
its spectrum is related to that of the triangular lattice through a simple
analytical transformation, which allows to compute the
eigenspectrum for rational values of the magnetic flux. When the magnetic
flux per rhombic plaquette equals half the flux quantum, one faces a rather
exotic effect for an infinite periodic tiling, since the density of states
degenerates into three $\delta $-peaks and the electron motion becomes
bounded.
We also present another example of an infinite periodic tiling denoted ${\cal T}_{4}$, which, while having a more complex
 geometry, displays the same properties as ${\cal T}_{3}$.\\
Other structures, like quasiperiodic or random tilings, can contain some bounded cages due to these particular star
configurations, but  generically, one does not expect a highly degenerated discrete spectrum as for ${\cal T}_{3}$ and ${\cal T}_{4}$.
In both cases, cages induce interesting
instances of wavepacket time evolution with well characterized pulsing states.
\\

Let us first examine ${\cal T}_{3}$, a bipartite periodic hexagonal structure with 3 sites per
unit cell (figure \ref{Star3lattice}), one 6-fold coordinated (denoted $A$) and two 3-fold
coordinated (denoted $B$ and $C$).
\noindent We consider an tight-binding hamiltonian defined by:
\begin{equation}
{\large {\cal H}}=\sum_{<i,j>}t_{ij}|i><j|
\label{hamil}
\end{equation}
where $t_{ij}=1$ if $i$ and $j$ are nearest neighbours, $0$ otherwise, and $%
|i>$ is a localized orbital on site $i$. In the presence of a magnetic field 
${\bf {H}}$, the hopping terms are modified by phase factors involving the
vector potential ${\bf {A}}$ \cite{Peierls}. Let $\gamma _{ij}$ be the phase
factor between sites $i$ and $j$: 
\begin{equation}
\gamma _{ij}={\frac{2\pi }{\phi _{0}}}\int_{i}^{j}{\bf A}.d{\bf l}
\end{equation}
where $\phi _{0}=hc/e$ is the flux quantum. We first consider a uniform magnetic field ${\bf H}=H{\bf z}$ in 
the Landau gauge: ${\bf A}=H(0,x,0)$.
As shown below, the whole spectrum only depends on the reduced flux $f=\phi
/\phi _{0}$ where $\phi =Ha^{2}\sqrt{3}/2$ is the magnetic flux through an
elementary rhombus. Taking into account the translational invariance in the $%
y$ direction due to the gauge choice, the eigenfunctions write:
\begin{equation}
\varphi (x,y)=\psi (x)e^{ik_{y}y},\hspace{1ex}k_{y}\in \left[
0,2\pi /{a\sqrt{3}}\right]
\end{equation}
which allows to map the spectrum onto the solution of an
effective one-dimensional problem: 
\begin{eqnarray}
(\varepsilon ^{2}-6)\psi _{m} &=&2\cos \left( {\frac{\gamma }{2}}\right) 
\left[ 2\cos \left( {\frac{3\gamma }{2}}\left( m+{\frac{1}{2}}\right)
+\kappa \right) \psi _{m+1}\right.   \nonumber \\
&& +\left. 2\cos \left( {\frac{3\gamma }{2}}\left( m-{\frac{1}{2}}\right)
+\kappa \right) \psi _{m-1} \right. \nonumber \\
&&+2\cos \left( 3\gamma m+2\kappa \right) \psi _{m} \bigg]
\label{spectre1d}
\end{eqnarray}
where $\psi _{m}=\psi _{A}(x)$ for $x=3ma/2$ $(m\in {\bf Z})$, is the wave
function at the $A$-type sites, $\kappa =k_{y}\sqrt{3}/2$, $\gamma =2\pi f$,
and $\varepsilon $ is the energy for the two-dimensional original problem.
Note that this mapping is only valid for $\varepsilon \neq 0$ (the $%
\varepsilon =0$ case is briefly adressed below). For the triangular lattice, a
similar equation has been derived by Claro and Wannier \cite{Claro79}: 
\begin{eqnarray}
\varepsilon _{T}\psi _{m} &=&2\cos \left( \gamma _{T}\left( m+{\frac{1}{2}}
\right) +\kappa \right) \psi _{m+1}  \nonumber \\
&&+2\cos \left( \gamma _{T}\left( m-{\frac{1}{2}}\right) +\kappa \right) \psi
_{m-1} \nonumber \\
&&+2\cos (2\gamma _{T}m+2\kappa )\psi _{m}
\end{eqnarray}
with similar notations. So, the ${\cal T}_{3}{\cal \ }$
spectrum can be obtained from the triangular lattice one by choosing 
$\gamma _{T}=3\gamma /2$; the eigenenergies of the two tilings being 
related by: 
\begin{equation}
\varepsilon ^{2}-6=2\cos \left( {\frac{\gamma }{2}}\right) \varepsilon _{T}
\label{homo}
\end{equation}
This extends a classical result on bipartite structures to the case of a
uniform magnetic field. Since the tiling is made up of rhombi, the spectrum
is symmetric ($\varepsilon \leftrightarrow -\varepsilon $). Moreover,
eq. ($\ref{spectre1d}$) displays a translation symmetry $f\rightarrow
f+n\;$$(n\in {\bf N})$ and a reflection invariance about half-integer values
of $f$. Thanks to these symmetries, we can limit our analysis to $0\leq
f\leq 1/2$. For rational values of $f=p/q$ ($p,q$ mutually prime), the
system ($\ref{spectre1d}$) becomes closed after a translation by $q$ periods
and the spectrum is made up of $2q$ bands. Note that, when $q=3q^{\prime }$ $%
(q^{\prime }\in {\bf N})$, this period is reduced by a factor $3$ (and the
spectrum has only $2q^{\prime }$ bands). An interesting example occurs for $%
f=1/3$ where the spectrum displays two symmetric bands extending from $\pm 3$
to $0$, and is therefore gapless.
In addition, for any value of the magnetic field, a close inspection of the secular equations system for
each type of site shows that the spectrum contains a highly degenerated eigenenergy (flat band) at 
$\varepsilon =0$ with weight $1/3$. Figure \ref{butterfly} shows the ${\cal T}_{3}$ spectrum
support versus $f$. The most remarkable point in this context, is that for $f=1/2$, the spectrum collapses into three
eigenvalues $\varepsilon =0$ and $\varepsilon ^{_{\pm }}=\pm \sqrt{6}$ (see eq. (\ref{homo})).\\

One way to analyze the link between the spectrum and the quantum dynamics is to study the
spectral properties at a local level. 
The magnetic field being uniform, the whole spectrum is recovered from
the three local density of states (LDOS); furthermore, by symmetry, the two
3-fold coordinated sites ($B$ and $C$) are equivalent, so that we just need
to study two different types of local environment. It is then suitable to
shift from the above Landau gauge to an equivalent cylindrical symmetric
gauge given by ${\bf A}=(H/2)(-y,x,0)$.\\
We proceed to an analytic Lanczos tridiagonalization
of local clusters, along the recursion method algorithm \cite{recursion}. An
effective semi-infinite chain is generated, with normalized orbitals $%
\left| n\right\rangle $ and hopping terms ($a_{n}$,$b_{n}$), according to
the three-term recurrence relation: 
\begin{equation}
b_{n+1}\left| n+1\right\rangle =({\cal H}-a_{n})\left| n\right\rangle
-b_{n}\left| n-1\right\rangle  \label{recurs}
\end{equation}
which allows to evaluate the LDOS at a given site (or any linear combination
of site orbitals). Since ${\cal H}$ is purely off-diagonal and ${\cal T}_{3}$
is bipartite (and therefore only contains even loops), the diagonal $a_{n}$
terms vanish ({\it i.e.} the odd moments). Notice that in this case, the orbital $%
\left| n\right\rangle $ encompasses all the sites reached from $%
\left| 0\right\rangle $ at least after  $n$ jumps, and can be interpreted in terms of shell orbitals.
If one of the $b_{n}$ vanishes, the effective chain 
becomes finite and the LDOS is discrete. It also implies that a wavepacket initially localized in the tiling
at the origin site will only spread over a finite set of sites belonging to the mutually coupled shells.\\ 

We first choose the initial orbital $
\left| 0\right\rangle $ at a 3-fold coordinated site and apply the recursion procedure. One then easily obtains:
 $b^3_{1}=\sqrt{3}, b^3_{2}=D_{2}/\sqrt{3}$ and $b^3_{3}=D_{3}/D_{2}$ where: 
\begin{eqnarray}
D_{2} &=&(12\cos ^{2}(\pi f)+9)^{1/2}  \nonumber \\
D_{3} &=&(24\cos ^{2}(\pi f)+12(\cos (\pi f)+\cos (3\pi f))^{2})^{1/2}.
\label{coefrec}
\end{eqnarray}
It is clear that $b^3_{3}=0$ for $f=1/2$, and that the LDOS
reduces to the following three eigenvalues $0,\pm \sqrt{6}$. A similar derivation for the
6-fold coordined site leads to: $b^6_{1}=\sqrt{6}, b^6_{2}=2\cos (\pi f)$. For $f=1/2$, $b^6_{2}$ also
vanishes and the LDOS reduces to $\pm \sqrt{6}$ (see figure \ref{recursion}).\\
We now proceed to describe the quantum dynamics on ${\cal T}_{3}$.
Owing to the superposition principle, we only consider initial distributions strictly localized
on each type site, given that any wavepacket behaviour can
be described in terms of these individual evolutions. For a generic value of 
$f$, the initially localized electron wavefunction spreads with time and
asymptotically vanishes at the origin, the time dependance of this phenomenon
being determined by the nature of the eigenspectrum. For $f=1/2$, the
quantum dynamics can be  derived in closed forms and it displays a very different
behaviour since the system remains localized inside a restricted region.
Indeed, an initial distribution localized on a $A$ site is completely ``trapped" inside its surrounding star
($b^6_2=0$) so that, at any time, there will only be
nonvanishing components of the wavefunction at the origin and on the first 
shell. This oscillating motion also occurs for a wavepacket initially localized on a $B$ (or $C$) site
 ($b^3_{3}=0$), but is here confined inside a larger neighborhood (see figure \ref{recursion}).
Consequently, for $f=1/2$, any wavepacket on ${\cal T}_{3}$ has a limited extent.\\
This extreme localization in a  magnetic field can be simply understood in
terms of Aharonov-Bohm interferences around the rhombi; we therefore call
the above confined regions $\cal{A.B.}$ cages. More precisely, for a given tiling embedded 
in a magnetic field ${\bf H}$, we associate to any site $\left| i\right\rangle $, 
an $\cal{A.B.}$ cage ${\cal C}^H_{i}$ defined as the set of sites reached
 by a wavepacket initially localized on $\left| i\right\rangle $.
Note that ${\cal C}^H_{i}$ is generically infinite, but specific values of the magnetic field \ ($f=1/2$ 
for ${\cal T}_{3}$) can leave it bounded. In this case, the cage can also
be characterized by its boundary sites, where completely destructive interferences
disconnect it from the rest of the tiling.
For a given structure, it is sufficient to characterize the inequivalent site
cages, and several cases can  occur:
\begin{itemize}
\item All  cages are unbounded for any $f$.\\
Ex: Square lattice, triangular lattice,
honeycomb.
\item Some cages are bounded for particular values of $f$.\\ 
Ex: Star patterns in Penrose tilings and octagonal tilings (see below).
\item For the same values of $f$, all cages 	are simultaneously bounded.
\end{itemize}

The simplest example of the latter case is encountered on  ${\cal T}_{3}$ for 
$f=1/2$, but it can also occur for other tilings. Let us pay attention to
${\cal T}_{4}$ represented figure in \ref{Star4lattice} which is an  approximant with $14$ sites per unit cell of the octagonal
quasiperiodic tiling.
This tiling contains three types of faces (large square, small square, and parallelogram), but only 
two different areas.\\ 
The previous recursion analysis can be done on the five different types of site  (A, B, C, D, E)
and one can prove that for $f=1/2$, all the site cages are bounded (in this case 
$f$ refers to the magnetic flux per small area). Thus, the eigenspectrum is also discrete and the quantum dynamics is
 similar to that discussed on 
 ${\cal T}_{3}$ (bounded extension of any wavepacket, pulsating states, ...). It is also possible to continuously 
deform ${\cal T}_{4}$, keeping the same connectivity, until all the edges have the same length, and the 
star around $A$ becomes 8-fold symmetric. But any metrical 
transformation changes the nature of the spectrum and the dynamics, so that only some site cages remains bounded 
(${\cal C}^H_{A}$ for $f=1/2$). In addition, when all the edge lengths are equal, the areas are incommensurate 
and the energy spectrum is no longer periodic with $f$. This loss of periodicity
does always occur when two tile areas are incommensurate as it is the case in quasiperiodic tilings. 
Indeed, consider the closed paths in the tiling contributing, for instance, to the fourth moment of the LDOS at
a generic site. Among them, only the non self-retracing ones are field
dependent, and lead to cosine terms whose arguments depend on the
circumnavigated rhombi areas. In those tilings, the spectrum becomes then
quasiperiodic with $f$ as remarked in \cite{Schwabe}\cite{Grimm}.\\
Concerning quasiperiodic tilings, some local configurations display interesting 
characteristics in the presence of a magnetic field. As for ${\cal T}_{3}$ and ${\cal T}_{4}$, 
star-like patterns can be encountered, providing canonical example of $\cal{A.B.}$ cages. An easy way
to understand the properties of such an environment is to use the symmetry group of the cluster. For convenience,
we only consider p-fold symmetric stars ($2p+1$ sites) (see figure \ref{recursion} (p=6)),
 whose first shell sites are 3-fold 
coordinated and where all the edges have the same length. In the absence of a magnetic flux, those stars
remains unchanged under $C_{pv}$ transformations, but the magnetic field breaks the reflection invariance. 
Finally, one  can only deal with the $C_{p}$ rotation group. We denote the $p$  one-dimensional irreductible
 representations by
$\Gamma _{i}$ $(i=0,p-1)$. If we consider the symmetrical representation $\Gamma _{0}$, one can show
that the LDOS at the star center reduces to $\varepsilon ^{_{\pm }}=\pm \sqrt{p}$, as soon as $f=1/2$. This result 
is nothing but a generalization of the above recursion study on the ${\cal T}_{3}$ $A$ site, to any $p$-order symmetry.
Consequently, the dynamical properties are also comparable and one faces  a simple example of $\cal{A.B.}$ cage
 that appears in quasiperiodic rhombus tilings (Penrose ($p=5,10$), octagonal ($p=8$)). A close 
investigation of the  other $p-1$ representations shows that when $f=m/p$ $(m=1,$ $ p-1)$, one can identify confined 
eigenstates associated to  $\varepsilon =0$, similar to those existing for $f=0$. 
Along the same line, one can prove that there exist strictly localized states in the clusters studied by Kohmoto 
{\it et al.} \cite{Kohmoto}, for any values of the magnetic field. Some of these states persist in the infinite 
tiling and have actually been numerically observed by Schwabe {\it et al.} \cite{Schwabe}.
\\
It is also important to clarify the effect of disorder on this field induced localization. Queerly, one expects
the randomness to alter and even destroy the phase matching essential for this localization effect. 
A standard way to disorder a system is to randomly modulate the hopping terms without changing the tiling configuration; 
but a different approach consist in flipping the rhombic tiles as in the Random Tiling Model \cite{Elser}. In  ${\cal T}_{3}$, a finite number of such flips should neither affect the discrete
nature of the spectrum nor the confined diffusion process. However, a finite density of flips may change those 
properties.\\
 
Finally, a natural question  arises about the possibility to experimentally observe this very strong localization
phenomenon. We should first remark that  $\cal{A.B.}$ cages are sensitive to various kinds of disorder which are likely
to occur in real systems: fluctuations in the tile areas and in the transmission matrix along the edges are 
strongly relevant, but unlikely to fully destroy the localized nature of the eigenstates. We may just conjecture a strange
 non-monotonous variation of the localization length as a function of  disorder strength.\\
A certainly interesting experiment would be to investigate the magnetic vortex lattice in a superconducting wire network
with  the  ${\cal T}_{3}$ geometry. Indeed, the unusual nature of the linearized Ginzburg-Landau equation eigenstates
is to be reflected on the behavior of the solutions in the presence of the non-linear term. A crucial issue would 
be to determine whether there is a unique  optimal lattice structure, or a glassy type of vortex-solid, or even a disordered
vortex-liquid. These various possibilities are still debated in simpler geometries \cite{f}. We believe that the 
systems presented here are good candidates for multiple ground state configurations, but the nature of the 
energy barriers connecting them is certainly a challenging open problem.

\acknowledgments 
 We would like to thank Cl. Aslangul, B. Delamotte, D. Mouhanna and B. Pannetier for fruitful discussions.

\vspace{10.ex}

{\bf Figure Captions}
\begin{enumerate}
\item{}
\label{Star3lattice}
A piece of ${\cal T}_{3}$ lattice embedded in a perpendicular magnetic field ${\bf {H}}$; 
$a$ is the rhombus edge length.

\item{}
\label{butterfly}
${\cal T}_{3}$ butterfly-like energy spectrum versus $f$.

\item{}
\label{recursion}
Effective semi-infinite chain (recursion framework), associated to  ${\cal T}_{3}$ local environments (6-fold site (left), 
3-fold site (right)).

\item{}
\label{Star4lattice}
A piece of the ${\cal T}_{4}$ periodic tiling. The five different types of sites are depicted\\
(A, B, C, D, E).

\end{enumerate}
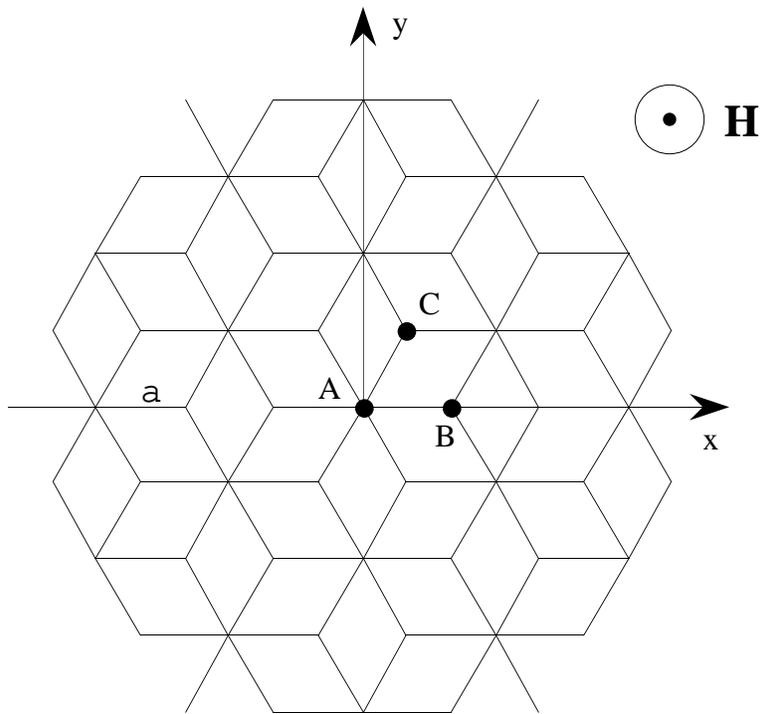
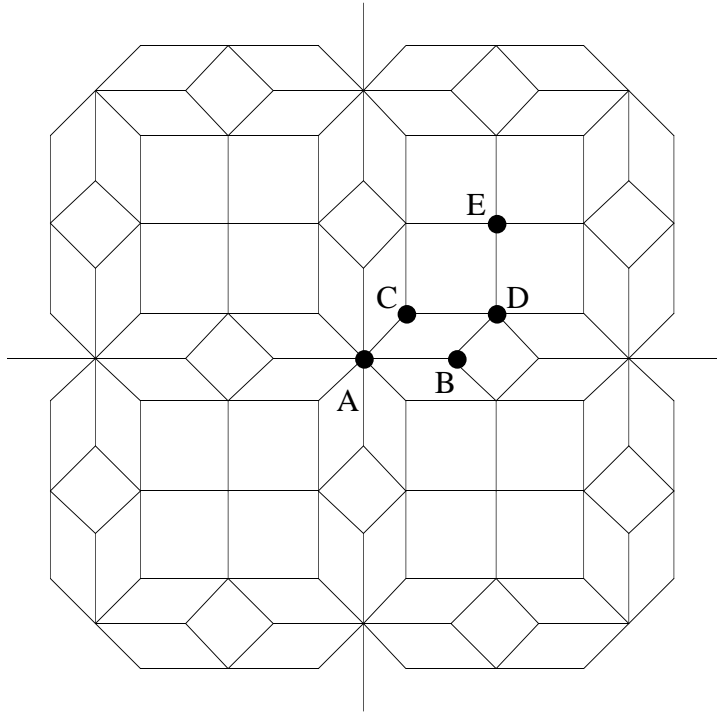
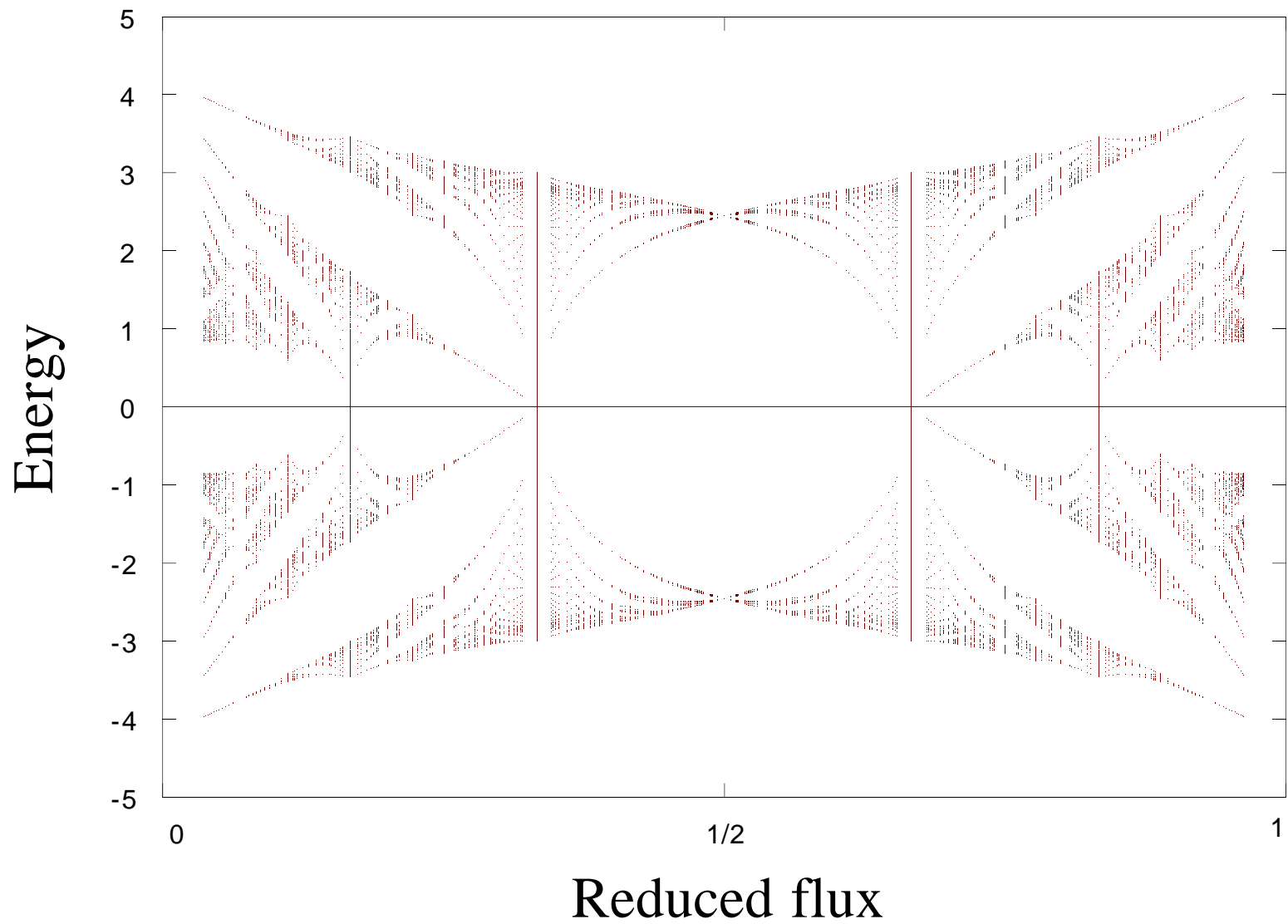
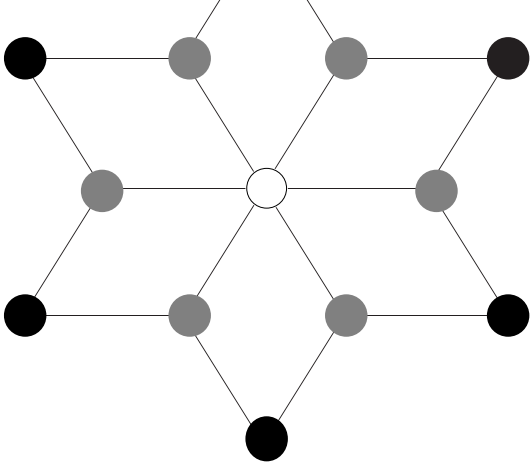
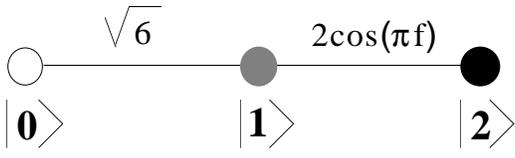
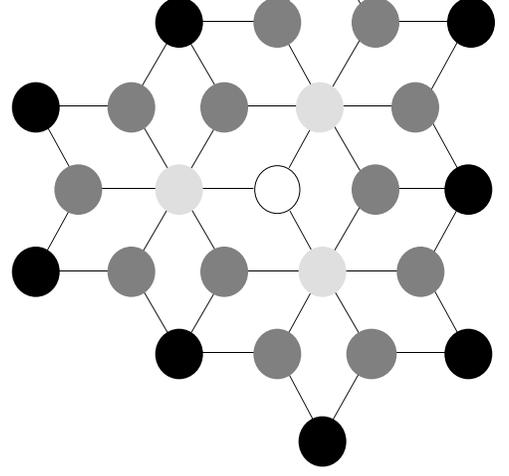
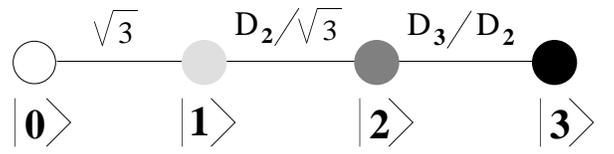
\end{document}